# APPLICATION OF FACEBOOK'S PROPHET ALGORITHM FOR SUCCESSFUL SALES FORECASTING BASED ON REAL-WORLD DATA

Emir Žunić[1,2], Kemal Korjenić[1], Kerim Hodžić[2,1] and Dženana Đonko[2]

[1]Info Studio d.o.o. Sarajevo, Bosnia and Herzegovina
[2]Faculty of Electrical Engineering, University of Sarajevo, Bosnia and Herzegovina

## ABSTRACT

*This paper presents a framework capable of accurately forecasting future sales in the retail industry and classifying the product portfolio according to the expected level of forecasting reliability. The proposed framework, that would be of great use for any company operating in the retail industry, is based on Facebook's Prophet algorithm and backtesting strategy. Real-world sales forecasting benchmark data obtained experimentally in a production environment in one of the biggest retail companies in Bosnia and Herzegovina is used to evaluate the framework and demonstrate its capabilities in a real-world use case scenario.*

## 1. INTRODUCTION

Generating product-level sales forecasts is a crucial factor in the retail industry since inventory control and production planning plays an important role in the competitiveness of any company that provides goods for its customers. While accurate and reliable forecasts can lead to huge savings and cost reductions by facilitating better production and inventory planning, competitive pricing and timely promotion planning, poor sales estimations are proven to be costly in this domain since it is well-known that goods shortages cause lower profits and can easily lead to customer dissatisfaction. Furthermore, not only the excess inventory may force the store to sell goods at lower prices, or even worse lead to inventory write-offs, higher than needed inventory levels also increase warehousing costs.

In the real-world scenario, the business environment in the retail industry is highly dynamic and often volatile, which is predominantly caused by holiday effects and competitor behaviour. As a result, contrary to the widely available academic datasets used to demonstrate and benchmark various time-series forecasting methods, real-world sales data in this domain carry various challenges, such as highly non-stationary historical data, irregular sales patterns, and highly intermittent sales data.

A module that would be able to forecast sales with a reasonably high accuracy, augmented by the module for highly-reliable classification of the product portfolio according to the expected level of forecastability, would be of great use for any company operating in the retail industry.

To bridge the gap towards the application of time-series forecasting in the real-world scenario in the retail industry, the focus of this work is set on the development of the module for reliable classification of the product portfolio according to the expected level of forecasting reliability. The results are presented on the example of dataset experimentally obtained in a production environment in one of the biggest retail companies in Bosnia and Herzegovina. Although generating sales forecasts anywhere between the daily and annual horizon is certainly possible, a particular focus has been put on monthly and quarterly sales forecasts since it was concluded from discussions with clients that the aforementioned period is of the greatest interest for production and inventory planning.

The structure of this paper is as follows: section Literature review offers a general description of previous studies relating to the use of the different approaches and algorithms in related retail sales forecasting problems, and methods for solving. Section Methodology gives an overview of the framework by briefly explaining a structure of input dataset, data filtering and preprocessing steps, the process of product portfolio selection, the Prophet tool, performance metrics and forecastability analysis using backtesting experiments, and guidelines for classifying the product portfolio, whereas section Results shows the capabilities of the proposed forecasting framework in a real-world use case scenario. The conclusions drawn from the results in terms of the proposed objective are given in section Conclusions, with a brief description of directions for future work, further development and application of the proposed framework.

## 2. LITERATURE REVIEW

To estimate future sales, the process of sales forecasting is used. Those accurate processes help companies to predict all kinds of performances and to make important business decisions. Company forecasts can be based on trends in economy, past sales data and comparisons in industry. Already established companies can easily predict future sales, which are based on past business data. New companies must create their forecasts on information, not being verified enough, such as competitive intelligence and market research. Sales forecasting enables approach into company's workforce, resources and cash flow. Predictive sales data is crucial for business in order to get investment capital.

Retail businesses must use their resources in an efficient way and make strategic decisions to make their revenues increased and stable, especially when conditions are getting more competitive. There are three main types of retail sales forecasting:

- Time-series sales forecasting,
- Sales forecasting based on Artificial Neural Networks,
- Using complex hybrid methods.

There are many studies in the literature with different simple and complex methods used for modelling sales data to forecast future sales.

In the paper by Aras *et al.* [1] the brilliant literature overview and comparative study on retail sales forecasting between single and combination methods is given. The obtained results in this paper suggested that the combination methods achieve better results than the individual ones. Also, the comparison between these methods and company's current system was done. Several other interesting facts are mentioned in this paper in the literature review section, as follows.

Sales data from the period of 10 years (from 1979 to 1989) were analysed by Ansuj *et al.* [2] in terms of ARIMA (Autoregressive Integrated Moving Averages) model with interventions and the ANN (Artificial Neural Network) model. Forecasts of the ANN model were more appropriate than the ARIMA ones. Comparative studies were made of traditional methods and ARIMA models with ANN models by Alon *et al.* [3], as well as a multivariate regression for aggregate retail sales in economic conditions that are stable and winter's exponential smoothing.

Results showed that the ANN models were the best. Frank *et al.* [4] created the women's apparel sales using the ANN model (which results were the best in terms of R2 evaluation statistics), Winter's three- parameter model and a single seasonal exponential smoothing. In sales forecasting, ANN models were the best of all. Aburto and Weber [5] created a replenishment system for a Chilean supermarket by using a hybrid methodology of two stages, whose forecasts were better in comparison to ANN and ARIMA models. There were not many sales failures in hybrid methodology, and there were lower inventory levels.

Au *et al.* [6] compared the performance of evolutionary neural networks for sales forecasting with the ARIMA seasonal model and totally connected neural network. Evolutionary neural networks produced more accurate forecasts. For forecasting retail sales, Pan *et al.* [7] suggested a hybrid method. That method integrates a neural network (EMD-NN) with an empirical mode decomposition to forecast retail sales. The conclusion was that the seasonal ARIMA model and the classical ANN model were less superior compared to the EMD-NN one. Performance with hybrid method is better in volatile economic conditions. In comparison made by Dwivedi *et al.* [8], the ANFIS (Adaptive Network-based Fuzzy Inference System) method was the most appropriate of all the other methods being compared, including ANN and linear regression and a neuro-fuzzy modelling approach. Aye *et al.* [9] made a performance out of 26 models (ANN, ARIMA, AFRIMA, etc.) in forecasting South Africa's aggregate seasonal retail sales. Results showed that nonlinear ANN model was outperformed by other models also being nonlinear.

In comparison of Ramos *et al.* [10], the results did not show any difference between the state space models and ARIMA model with automatic algorithms in forecasting sales of women's footwear products. Fabianová *et al.* [11] made an analysis of refrigerator sales from a retail store. Results were better in achieving the total revenue by Monte Carlo simulation and using sensitivity analysis for variables identification. Kolassa [12] observed forecast accuracy measures not being appropriate for count data. He took into consideration discrete predictive distributions for forecasting daily sales. Ma *et al.* [13] presented the results of examination of the case of Stock Keeping Unit (SKU) level retails store sales by using a four step methodological framework. This research showed that improvements were achieved by exploiting the intra red category information, not the inner one. Sales forecasting detailed review was provided in this way. Jiménez *et al.* [14] wanted to have forecasts for online sales being more accurate and also the relevant features of the solid products affecting the sales, so he proposed a selection methodology of a novel feature.

Retail product sales data are contained of multiple seasonal cycles of different lengths. For example, beer daily sales data shown in one experiment exhibit both weekly and annual cycles. Sales are high during the weekends and low during the weekdays, high in summer and low in winter, and high around Christmas. Some sales data depend on the nature of the business and business locations. So, models used in forecasting must control multiple seasonal patterns. Ramos and Fildes [15] use models with additional flexibility but parsimonious complexity to capture the seasonality of weekly retail data: trigonometric functions prove sufficient. Papacharalampous and Tyralis [16] consider the performance of random forests and Facebook's Prophet in forecasting daily streamflow up to seven days ahead in a river in the US. Both these forecasting methods use past streamflow observations, while random forests additionally use past precipitation information. They use a naïve method based on the previous streamflow observation, as well as a multiple linear regression model utilizing the same information as random forests. The obtained results suggest that random forests perform better in general terms, while Prophet outperforms the naïve method for forecast horizons longer than three days. Based on the knowledge about the sales forecasting, it is worth mentioning that we are currently working on integrating this framework into our previously developed products for the retail industry as a big Smart supply chain management (SCM) concept (Zunic *et al.* [17, 18, 19]). Also, the proposed forecasting method together with the global positioning system (GPS) data could be successfully used to determine some parameters and constants of the real-world vehicle routing problems (VRP), such as unloading time, road and time distances between customers and so on (Zunic *et al.* [20, 21, 22]).

# 3. METHODOLOGY

A module that would be able to forecast sales with a reasonably high accuracy, augmented by the module for highly-reliable classification of the product portfolio according to the expected level of forecastability, would be of great use for any company operating in the retail industry. The proposed model for successful sales forecasting based on real-world data is shown in Figure 1.

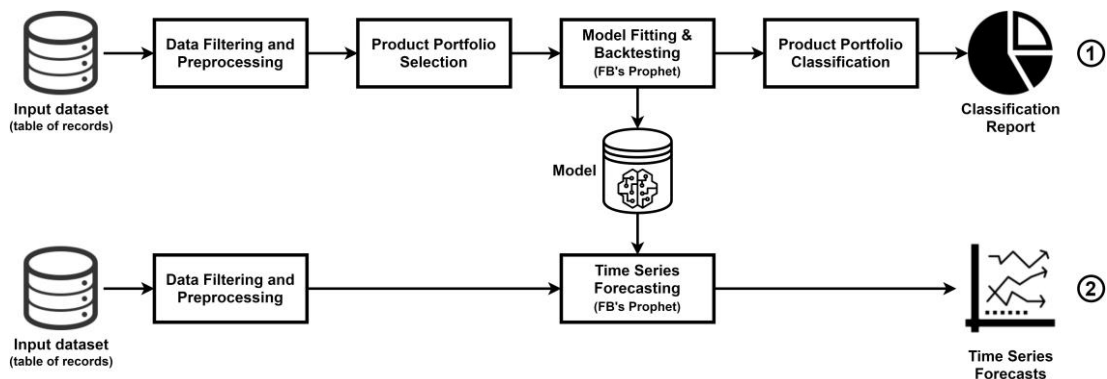

Figure 1. Proposed sales forecasting model.

The upper part (1) in the illustrated model can be represented as the "offline" segment in the whole approach, for purposes of model accuracy and classification. The second part (2) represents the sub-process of successful sales forecasting.

## 3.1. Input Dataset

To develop a framework for sales forecasting, the following columns were assumed to be available in the real-world input dataset which is structured as a table of records:

1) item_code - unique identifier of the product in a portfolio
2) date - date of transaction
3) quantity - the quantity sold in a given transaction
4) unit_price - the unit price at which the product was sold (optional, not used in forecasting)

A sample of input dataset is shown in Table 1.

Table 1. A sample of the input dataset

|   | item_code | date | quantity | unit_price |
|---|---|---|---|---|
| 0 | 501001000001 | 2010-01-02 | 399 | 1.3300 |
| 1 | 501001000001 | 2010-01-04 | 812 | 1.3380 |
| 2 | 501001000001 | 2010-01-05 | 516 | 1.3310 |

## 3.2. Data Filtering and Preprocessing

Several steps were taken during the preprocessing phase to transform a table of records into a convenient form:

— filtering by date, in order to remove irrelevant historical data (e.g. it was decided not to use more than six years of historical data)

— conversion of quantities into the same unit (e.g. pieces, packs, bundles, pallets, etc.)

— data aggregation in time domain at the product level (i.e. daily sales data were converted into monthly sales)

This process is illustrated in Figure 2.

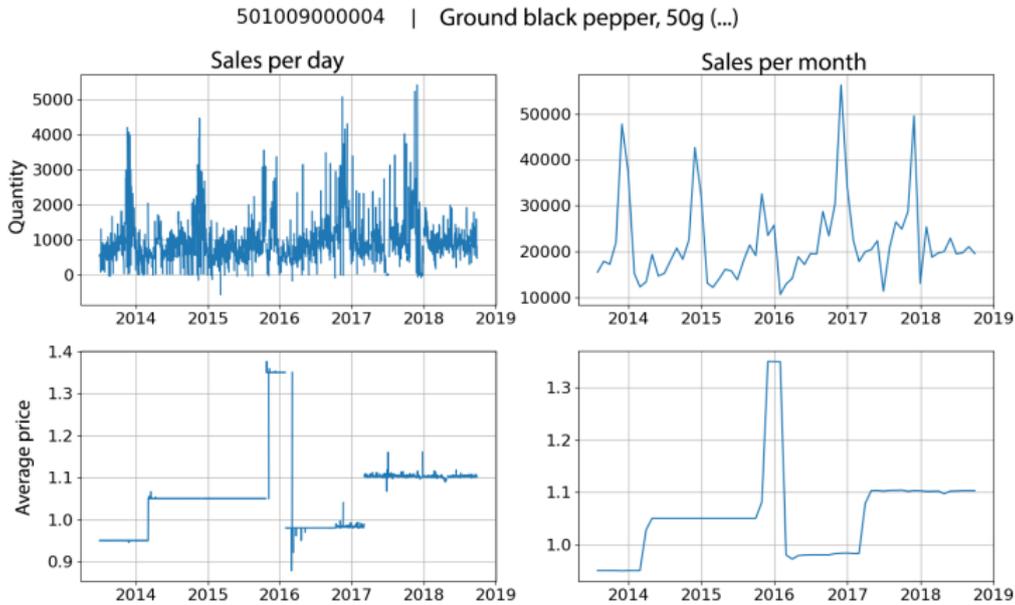

Figure 2. The illustration of data filtering and preprocessing step for a single item in the product portfolio

### 3.3. Product Portfolio Selection

To simplify the analysis by limiting it to a reasonable number of products from the portfolio, but at the same time properly deal with a long tail phenomenon present in sales data of any company in the retail industry, the products are first sorted by their importance.

Based on industry experience and discussion with clients in real field, it was concluded that several different criteria can be used to sort the product portfolio by relevance:

1) the total profit per product over the last year (if the profit per sale is available, which is a rarity),

2) the total financial turnover (i.e. net sales) per product over the last year (if the unit price per sale is available, which is often a case),

3) the total quantity sold per item over the last year (if none of the aforementioned data is available).

One can conclude that the second criterion is a quite good approximation of the first one since the percentage profit per product is usually comparable across the product portfolio. On the other side, the third criterion is a fairly loose approximation since the unit price per item may vary significantly, but it is the best that can be done if the price data is not available.

Afterwards, to perform forecastability analysis and product portfolio classification, as well as to present capabilities of sales forecasting framework and initial results to the client, the focus is set on the Top N products that cover 90% of the total profit/turnover/quantity over the last year. In a practical application of sales forecasting framework, after obtaining initial results in this way, one can easily re-run the experiment for the entire product portfolio and deploy the model. In the next step, products that are not suitable for product portfolio classification framework are filtered out according to the following requirements:

– minimum length of observation horizon: 39 months (at least 24 months of historical data is required for reliable estimation of trend and/or seasonal effects, additional 12 months of data for repeated backtesting experiments and three months of data to measure the accuracy of quarterly forecasts),
– maximum allowed production/sales downtime: 3 months (the product is considered to be inactive if zero sales are recorded for more than 3 months of the most recent history).

Since there is usually a non-negligible number of products with a historical data available for more than 24 months but less than 39 months (i.e. products for which forecasts can be generated, but backtesting cannot be done as described in the following subsection since there is no enough historical data available), it is desirable to explain to clients that forecasting is still possible but reliable estimation of forecasting accuracy cannot be calculated for these products. In that case, it is suggested to use sales forecasting results in a semi-automated manner, (i.e. with a human-in-the-loop).

## 3.4. Prophet - a Tool for Time Series Forecasting at Scale

The basic building block of the proposed framework for sales forecasting and product portfolio classification is a tool/method for generating high-quality time-series forecasts. Despite the fact that there are numerous tools/methods that can be applied, it was decided to use Facebook's Prophet tool for this research since it is capable of generating forecasts of a reasonable quality at scale.

Prophet, an open-source software released by Facebook's Core Data Science team, is a procedure developed for forecasting time series data based on an additive model where non-linear trends are fit with yearly, weekly, and daily seasonality, plus holiday effects. It works best with time series that have strong seasonal effects and several seasons of historical data. Prophet is robust to missing data and shifts in the trend, and typically handles outliers well. According to Taylor and Letham [23] research, Prophet is used in many applications across Facebook for producing reliable forecasts and performs better than any other approach in the majority of cases.

In this paper, Facebook's Prophet tool is used for modelling the dynamics of sales for items in a product portfolio without using additional regressors, with the aim of generating monthly and quarterly sales forecasts. It is worth mentioning that an empirical method for tweaking model parameters is used to incorporate domain knowledge into the proposed framework, but the same parameters are used for the entire product portfolio to avoid overfitting. It is empirically concluded that at least 24 months of historical data is required for reliable estimation of trend and/or seasonal effects. An example of using Facebook's Prophet tool to forecast the future sales for the product with sales per month time series is shown in Figure 2 for the next three months is illustrated in Figure 3.

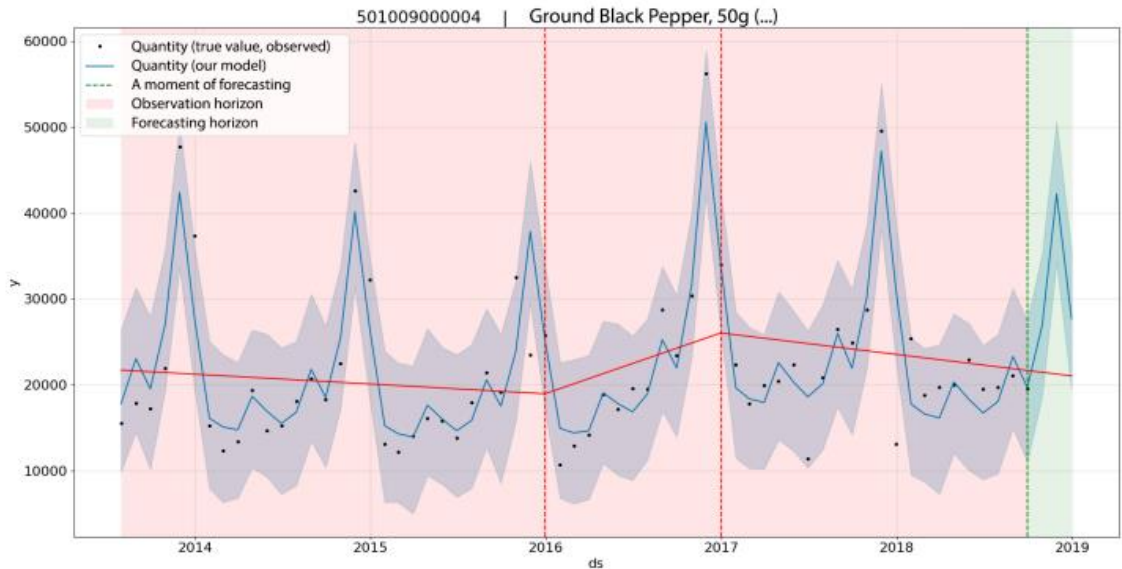

Figure 3. The illustration of using Prophet tool to forecast sales of the product with sales per month time series illustrated in Figure 2

## 3.5. Performance Metrics

Two performance metrics are used for measuring forecasting accuracy, calculating the expected level of forecasting accuracy and classifying product portfolio accordingly:

− the relative or percentage error (*PE*) for individual monthly/quarterly forecasts, and
− the mean absolute percentage error (*MAPE*) for quantifying the overall accuracy.

The percentage error (*PE*), which can be calculated as:

$$PE = \frac{y_{forecast} - y_{true}}{y_{true}} \cdot 100\% \qquad (1)$$

is mainly used to measure the accuracy of individual monthly/quarterly forecasting outputs generated by the model, while the mean absolute percentage error (*MAPE*), calculated as:

$$MAPE = \frac{1}{n} \sum_{i=1}^{n} \left| \frac{y^i_{forecast} - y^i_{true}}{y^i_{true}} \right| \cdot 100\% \qquad (2)$$

is used to quantify the overall accuracy of the forecasting framework and calculate the expected level of reliability useful for classifying the product portfolio.

These metrics are selected for use because of their simplicity, very intuitive interpretation, as well as the fact they work well if there are no extremes in the data. Based on the interaction with clients, it is concluded that a clear and intuitive interpretation of forecasting accuracy metrics in terms of relative error plays a crucial role in the acceptance of the sales forecasting framework as a decision-making tool in the retail industry.

### 3.6. Forecastability Analysis using Backtesting Experiments

To calculate a reliable estimation of the expected level of forecasting accuracy in terms of percentage error, an expanding window backtesting strategy is implemented. Past conditions are simulated by setting the present moment anywhere in the past, building a model using historical data to forecast future sales and see how accurately it would have predicted actual data.

One can assume that, with enough repetitions of simulating past conditions, a highly-reliable estimation of the expected level of forecastability can be calculated without having to wait for a new event to happen in order to compare it with previously generated forecasts and draw conclusions. To take into account yearly seasonality effects, it is recommended to carry out at least 12 repetitions of backtesting experiment with one month step size should be done, which is the reason why the limit for the minimum length of observation horizon is set to 39 months (i.e. at least 24 months of historical data is required for reliable estimation of trend and/or seasonal effects, 12 months for repeated backtesting experiments and three months of data to measure accuracy of quarterly forecasts).

In this research, the backtesting experiment is repeated 12 times with one-month step size for items selected from the product portfolio. At each step, a Prophet model is fitted to the data from the historical data (i.e. observation horizon) and monthly sales forecast for the next three months (i.e. forecasting horizon) are compared with the observed sales for the same period in order to calculate percentage error (*PE*) for monthly and quarterly forecasts. Then, the expected level of forecastability is calculated as the mean absolute percentage error (*MAPE*) over repeated backtesting experiments, which is used to quantify the expected level of forecasting reliability.

An example of performing forecastability analysis for a single item in the product portfolio is illustrated in Figures 4 and 5.

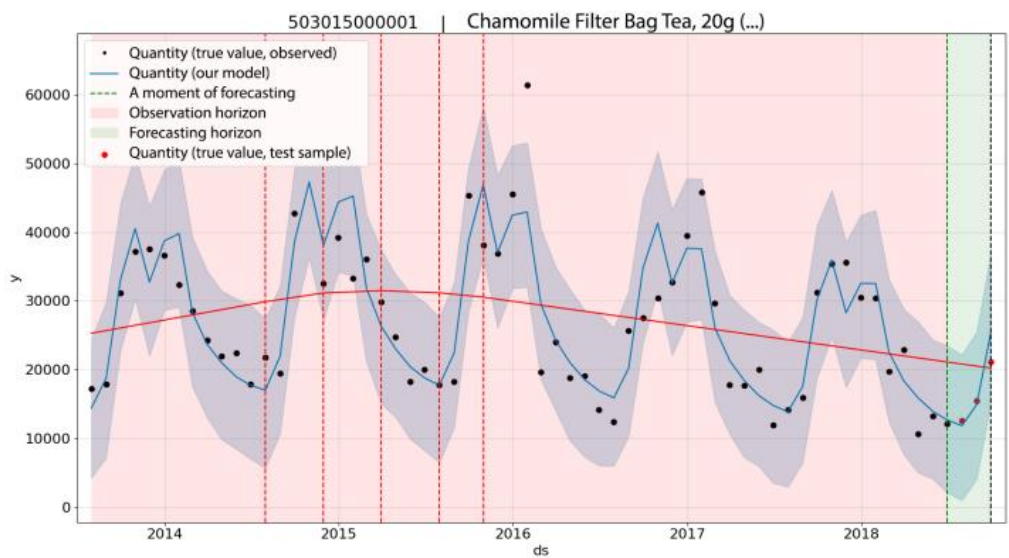

Figure 4. Step 1/12 of backtesting experiment illustrated on an item in a product portfolio

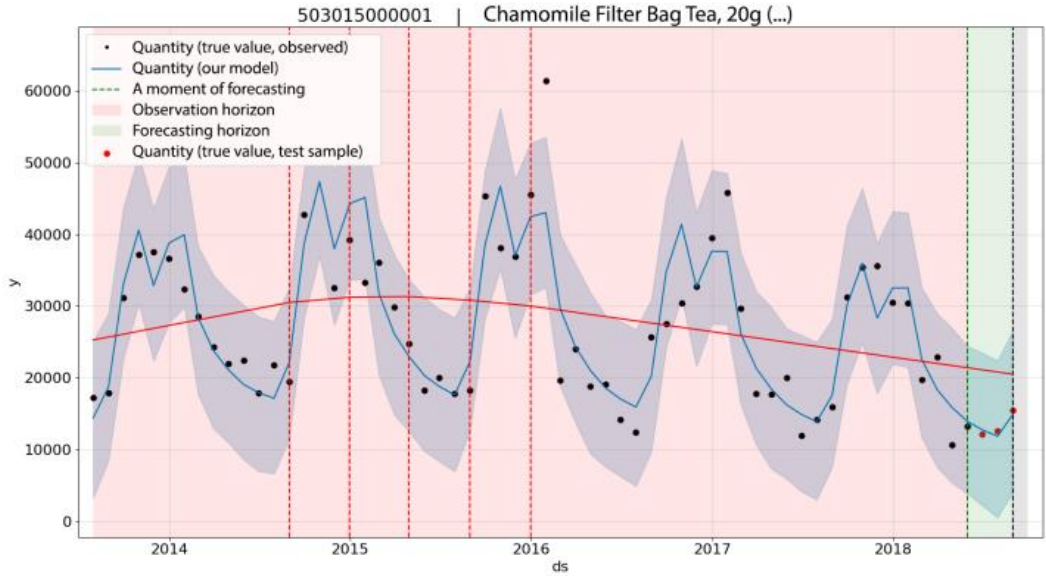

Figure 5. Step 2/12 of backtesting experiment illustrated on an item in a product portfolio

For the analysed item (Chamomile Filter Tea, 200g) the monthly mean absolute percentage error (*MAPE*) is:

$$MAPE = \frac{1}{n}\sum_{i=1}^{n}\left|\frac{y^{i}_{forecast} - y^{i}_{true}}{y^{i}_{true}}\right| \cdot 100\% = 8.00\% \qquad (3)$$

The quarterly forecast *MAPE* for the same item is:

$$MAPE = \frac{1}{n}\sum_{i=1}^{n}\left|\frac{y^{i}_{forecast} - y^{i}_{true}}{y^{i}_{true}}\right| \cdot 100\% = 6.06\% \qquad (4)$$

The obtained results for all the analysed items (real-world usa case scenario) are presented in the next section.

## 4. RESULTS

The proposed sales forecasting and product portfolio classification framework is evaluated in a real-world use case scenario with Real-world sales forecasting benchmark data published by Žunić [24], which is obtained experimentally in a production environment in one of the biggest retail companies in Bosnia and Herzegovina. Dataset is placed on 4TU.ResearchData in order to be available to the rest of the researchers, as a new benchmark data.

In this dataset, a total number of 581 items in product portfolio are observed, while 400 of these were active (i.e. non-zero sales are observed at least once) over the past year.

According to the guidelines proposed in the previous section, items in a product portfolio are ordered by the total financial turnover (i.e. net sales) per product over the last year and Top 200 items are selected with the aim of covering approximately 90% of total financial turnover.

## 4.1. Product Portfolio Classification

The first criterion used to classify the product portfolio is based on the observation horizon length and the recent sales downtime, so the following categories might be identified:

1) Inactive products: a subset of the product portfolio with items that have recent sales downtime of three or more months
2) Products with observation horizon shorter than 24 months: a subset of the product portfolio with items that cannot be forecasted nor classified according to the expected level of forecastability
3) Products with observation horizon shorter than 39 months: a subset of the product portfolio with items that can be forecasted but cannot be classified according to the expected level of forecastability
4) Products with observation horizon at least 39 months long: a subset of the product portfolio with items that can be both forecasted classified according to the expected level of forecastability

Classification of Top 200 products in a portfolio by this criterion is illustrated in Figure 6.

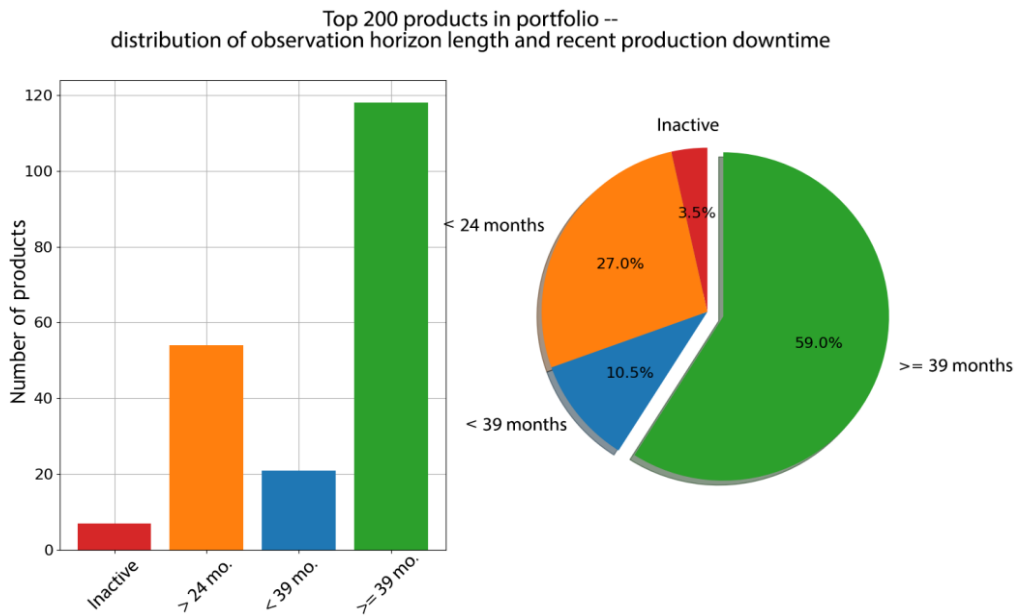

Figure 6. Classification of Top 200 products by the criterion based on the observation horizon length and the recent sales downtime

The second criterion for classifying product portfolio, which can be applied to a subset of products with observation horizon at least 39 months long, is based on the expected level of forecasting accuracy calculated as mean absolute percentage error (*MAPE*) for repeated backtesting experiments.

For example, binning products into the following class intervals might be interesting:

1) $MAPE \leq 15\%$
2) $15\% < MAPE \leq 30\%$
3) $30\% < MAPE \leq 50\%$
4) $MAPE > 50\%$

Classification of 113/200 products with observation horizon at least 39 months long (from the list of Top 200 products in a portfolio) by this criterion both for monthly and quarterly forecasts is illustrated in Figures 7 and 8.

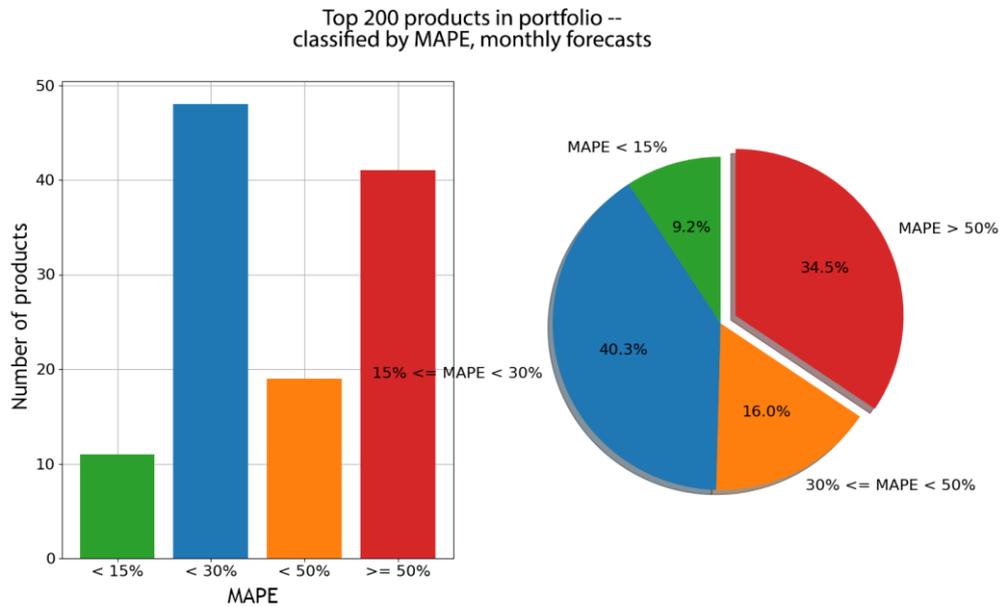

Figure 7. Classification of 113/200 products with observation horizon at least 39 months long: *MAPE* of monthly forecasts

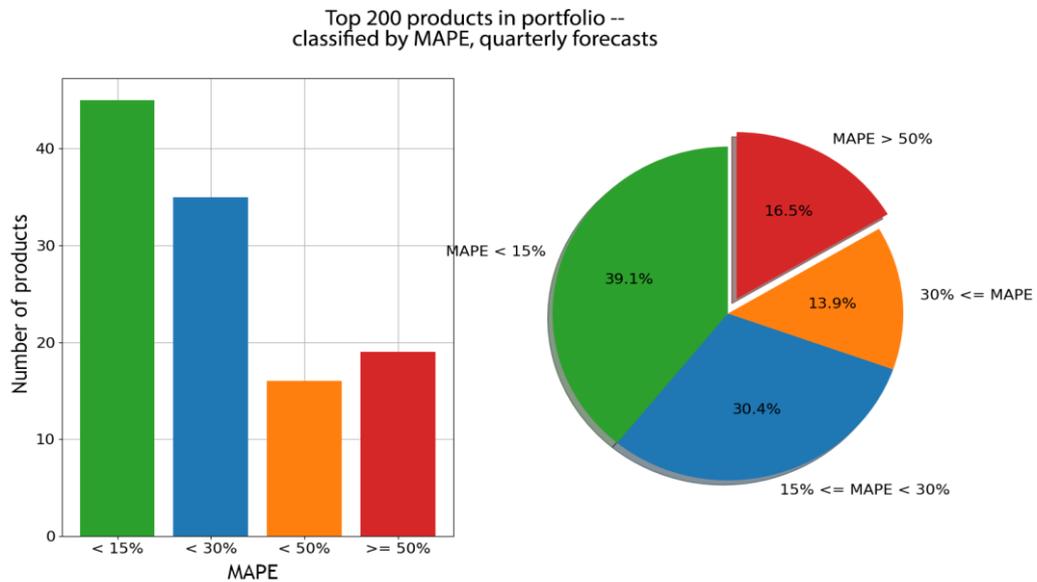

Figure 8. Classification of 113/200 products with observation horizon at least 39 months long: *MAPE* of quarterly forecasts

## 5. CONCLUSIONS AND FUTURE WORK

By evaluating its performance in a real-world use case scenario, the proposed framework demonstrated capabilities of generating reasonably accurate monthly and quarterly sales forecasts, as well as a great potential for classification of the product portfolio into several categories according to the expected level of forecasting reliability: approximately 50% of the product portfolio (with a sufficiently long historical data) can be forecasted with $MAPE < 30\%$ on a monthly basis, while approximately 70% can be forecasted with $MAPE < 30\%$ on a quarterly basis (40% of which with $MAPE < 15\%$).

It is important to mention that these approximately 40% of the product portfolio which can be forecasted with $MAPE < 15\%$ on a quarterly basis are mostly the best selling items of the aforementioned retail company, with more than 80% of annual share (financial) in the whole portfolio. Based on those facts, the obtained results are more than satisfactory in real-world scenario of sales forecasting.

With the aim of expanding a set of products to which the proposed framework can be applied, future work will also include the development of appropriate sales forecasting method applicable to products with observation horizon shorted than 24 months. This group of products represent a non-negligible part of the product portfolio, which is a major limitation of the proposed framework that will be tacked in the future. Further development of the proposed framework may include an automated approach for hyper-parameters tuning and optimization, modelling the impact of price changes, promotional activities and changes in the product portfolio, integrating multiple forecasting tools besides Prophet (such as X-13ARIMA-SEATS) and addressing some of its limitations described in this paper.

## ACKNOWLEDGEMENTS

The authors want to thank the company "Info Studio d.o.o." from Sarajevo, Bosnia and Herzegovina, for making this research possible through funding and providing access to necessary data.

## AUTHORS

**Emir Žunić** is a PhD candidate in Electrical Engineering with over 10 years of experience in the fields of Software Engineering, IT, Data Mining, Business Process Management, Document Management and Optimizations. He currently works as the Head of AI/ML Department at Info Studio d.o.o. Sarajevo. Also, he is the Co-Founder and CIO of edu720 d.o.o. Sarajevo. In the Academic Area, he also has experience in working as a Teaching Assistant/Industry Expert at the Faculty of Electrical Engineering, University of Sarajevo. In the past, he also worked as an Industry Expert at the Sarajevo School of Science and Technology. He has currently 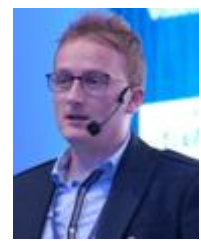 published 44 scientific papers at prestigious conferences and journals. He is an Editorial Board Member on several scientific conferences and journals.

**Kemal Korjenić** is a research engineer and data scientist at Info Studio d.o.o. Sarajevo. He gained his bachelor and master degree at the Faculty of Electrical Engineering, University of Sarajevo.

**Kerim Hodžić** is a Teaching Assistant at the Faculty of Electrical Engineering, University of Sarajevo and a research engineer at Info Studio d.o.o. Sarajevo. He gained his bachelor and master degree at the Faculty of Electrical Engineering, University of Sarajevo. He is a PhD student at the Faculty of Electrical Engineering, University of Sarajevo.

**Dženana Đonko** is a Full Professor at Faculty of Electrical Engineering, University of Sarajevo with enormous experience in the fields of data mining, machine learning and software engineering. She has published more than 50 scientific papers at prestigious conferences and journals. She is an Editorial Board Member on several scientific conferences and journals.